\documentclass[amsmath,amssymb,aps,pra,preprint,groupedaddress]{revtex4-2}

\usepackage[dvipdfmx]{graphicx}
\usepackage{dcolumn}
\usepackage{bm}
\usepackage{longtable}
\usepackage{ulem}
\usepackage{xcolor,soul}

\newcommand*\Laplace{\mathop{}\!\mathbin\bigtriangleup}

\begin{document}


\title{Langevin Navier-Stokes simulation of  protoplasmic streaming by 2D MAC method  }


\author{Shuta Noro \textit{$^{1}$}}
\author{Satoshi Hongo \textit{$^{1}$}}
\author{Shinichiro Nagahiro \textit{$^{1}$}}
\author{Hisatoshi Ikai \textit{$^{1}$}}
\author{Hiroshi Koibuchi \textit{$^{2}$}}\email[]{koibuchi@ibaraki-ct.ac.jp; koibuchih@gmail.com}
\author{Madoka Nakayama \textit{$^{3}$}}
\author{Tetsuya Uchimoto \textit{$^{4}$}}
\author{Jean-Paul Rieu \textit{$^{5}$}}
\affiliation{
$^{1}$\quad~National Institute of Technology (KOSEN), Sendai College, 48 Nodayama, Medeshima-Shiote, Natori-shi, Miyagi 981-1239, Japan \\
$^{2}$\quad~National Institute of Technology (KOSEN), Ibaraki College, 866 Nakane, Hitachinaka, Ibaraki  312-8508, Japan \\
$^{3}$\quad~Research Center of Mathematics for Social Creativity, Research Institute for Electronic Science, Hokkaido University, Sapporo, Japan \\
$^{4}$\quad~Institute of Fluid Science (IFS), Tohoku University, 2-1-1 Katahira, Aoba-ku Sendai 980-8577, Japan \\
$^{5}$  Univ Lyon, Universit$\acute{e}$ Claude Bernard Lyon 1, CNRS UMR-5306, Institut Lumi$\grave{e}$re Mati$\grave{e}$re, F-69622, Villeurbanne, France
}




\begin{abstract}
We study protoplasmic streaming in plant cells such as chara brauni by simplifying the flow field to a two-dimensional Couette flow with Brownian random motion inside parallel plates. Protoplasmic streaming is receiving a lot of attention in many areas, such as agriculture-technology and biotechnology. The plant size depends on the velocity of streaming and the driving force originating in molecular motors. Therefore, it is interesting to study detailed information on the velocity of streaming. Recently, experimentally observed peaks in the velocity distribution have been simulated by a 2D Langevin Navier-Stokes (LNS) equation for vortex and flow function. However, to simulate actual 3D flows, we have to use the NS equation for velocity, which, in the case of 2D flows, is not always equivalent to that for vorticity and stream function. In this paper, we report that a 2D LNS equation for velocity and pressure successfully simulates protoplasmic streaming by comparing the results with the experimental data and those obtained by 2D LNS simulations for vortex and flow function. Moreover, a dimensional analysis clarifies the dependence of numerical results on the strength $D$ of Brownian random force and physical parameters such as kinematic viscosity and cell size. We find from this analysis how the peak position in normalized velocity distribution moves depending on these  parameters.
\end{abstract}

\maketitle


\section{Introduction}
\begin{figure}[!h]
\centering{}\includegraphics[width=8cm]{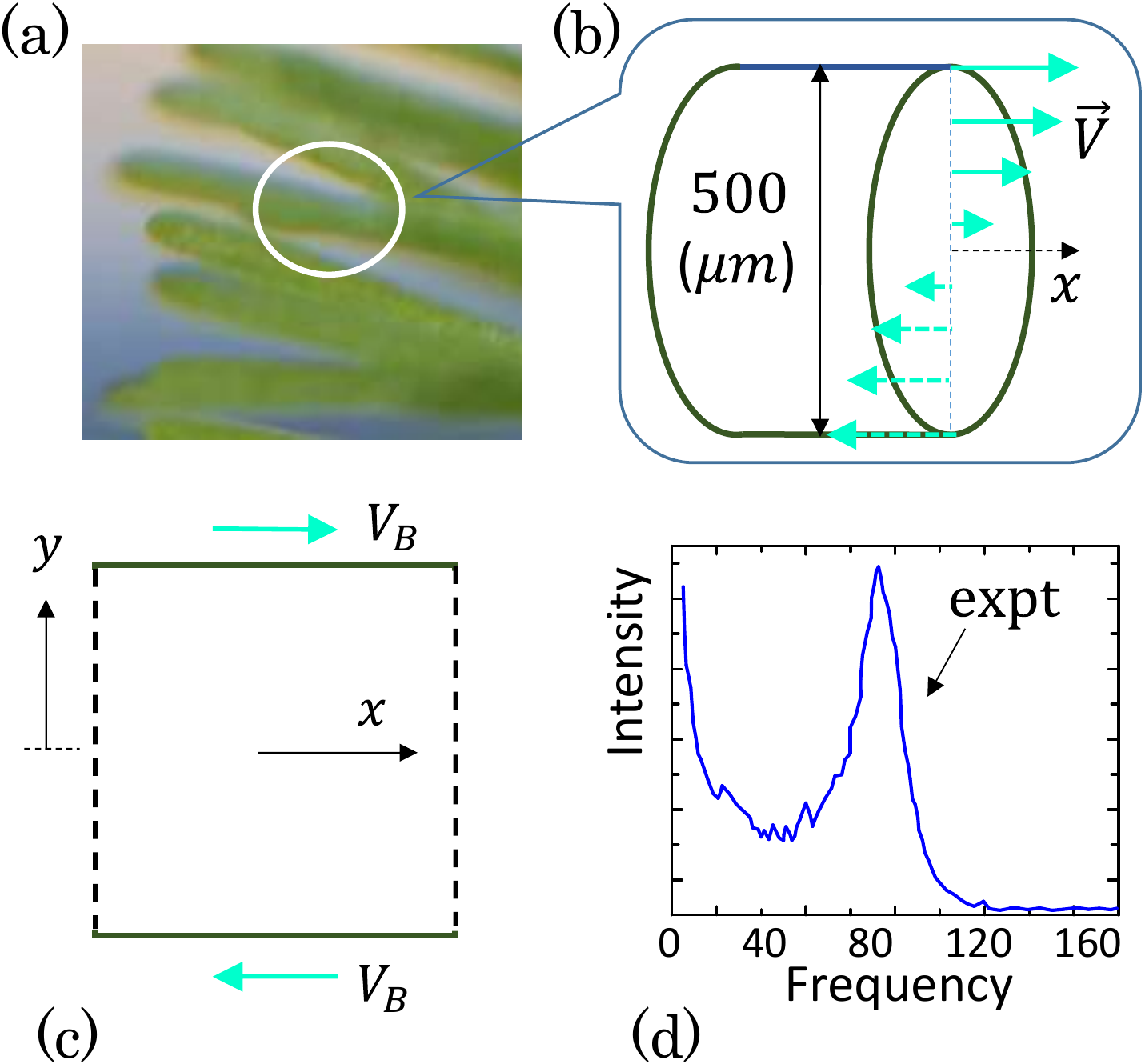}
\caption{(a) A photograph of plant in water, (b) illustration of protoplasmic streaming, (c) 2D computational domain, and (d) experimentally observed peaks in the velocity distribution $h(V)$ in Ref.~\cite{Mustacich-Ware-BJ1976}. 
Arrows in (b) denote velocity of flows, where circular flows on the cylindrical surface in real plant cells are modified to be parallel to the $x$ direction, and $V_B$ in (c) denotes the fixed boundary velocity.  }
\label{fig-1}
\end{figure}

 Protoplasmic streaming is a fluid flow directly observable in plants in water such as chara corallina~\cite{ShimmenYokota-COCB2004,VLubics-Goldstein-Protoplasma2009} (Figure \ref{fig-1}(a) shows a plant in water).   Kamiya and Kuroda reported the position dependence of the streaming in Nitella cells by a microscope~\cite{KamiyaKuroda-BMT1956}. The cell's typical diameter size is $500 ({\rm \mu m})$, and the maximum velocity is about $50 ({\rm \mu m/s})$  (Fig. \ref{fig-1}(b)). The kinematic viscosity is known to be approximately $\nu\!=\!1\!\times\!10^{-4} ({\rm \mu m^2/s})$, which is 100 times larger than that of water~\cite{KamiyaKuroda-BMT1973}. Recently, Tominaga and Ito reported that the flow speed influences the plant size~\cite{TominagaIto-COPB2015}. In addition, the activation force of the streaming is at present known to be the so-called molecular motor, and this mechanism of activation is the same as in the case of animal cells~\cite{McintoshOstap-CSatAG2016,Astumian-Science2020,Julicher-etal-RMP1997}. For these reasons, protoplasmic streaming has attracted much attention as a microfluidic flow in nanotechnological areas~\cite{Squires-Quake-RMP2005}. 

%
 
 Meent et al. studied the flow field both theoretically and experimentally~\cite{Goldstein-etal-PRL2008,Goldstaein-etal-PNAS2008,Goldstaein-etal-JFM2010,Raymond-Goldstein-IF2015}, and  confirmed that the reported data in Ref.~\cite{ KamiyaKuroda-BMT1956} are almost correct. A particle image velocimetry technique has been confirmed to be efficient for the streaming~\cite{Kikuchi-Mochizuki-PlosOne2015}. Niwayama et al. numerically studied the position dependence of velocity in plant cells~\cite{Niwayama-etal-PNAS2010}, and the results are consistent with those in Refs.~\cite{Goldstein-etal-PRL2008,Goldstaein-etal-PNAS2008,Goldstaein-etal-JFM2010,Raymond-Goldstein-IF2015}. 
 
 About twenty years after the report of Ref.~\cite{KamiyaKuroda-BMT1956}, Mustacich et al. experimentally observed the velocity distribution by a technique called Laser Doppler velocimetry and found that there appear two different peaks in the distribution at $V_x\!=\!0$  (or $V\!\to\!0$ )  and $V\!\not=\!0$~\cite{Mustacich-Ware-PRL1974,Mustacich-Ware-BJ1976,Mustacich-Ware-BJ1977,Sattelle-Buchan-JCS1976}. The peak at  $V\!\to\!0$ corresponds to the   zero-velocity Brownian motion of fluid particles, while the peak at $V\!\not=\!0$ is considered to correspond to the speed of  the molecular motor.  Recently, these two different peaks have been numerically studied by Langevin Navier-Stokes (LNS) simulations with vortex ($\omega$) and stream function ($\psi$)  for two-dimensional (2D) Couette flow between parallel plates~\cite{Egorov-etal-POF2020}. 
 
 However, the technique in Ref.~\cite{Egorov-etal-POF2020} is limited to only  2D flows because the NS equation is for $\omega$ and $\psi$  and is not applicable to actual 3D flows. Therefore, studying 2D Couette flow by LNS simulations for velocity ($\vec{V}$) and pressure ($p$) is worthwhile. 
 
 In this paper,  we study the velocity distribution of 2D Couette flow by simulating the LNS equation for $\vec{V}$ and $p$  and compare the results obtained in Ref.~\cite{Egorov-etal-POF2020} and experimental data in Refs.~\cite{Mustacich-Ware-PRL1974,Mustacich-Ware-BJ1976,Mustacich-Ware-BJ1977,Sattelle-Buchan-JCS1976}. Dimensional analysis of the LNS equation for $\vec{V}$ and $p$ is also performed, and we obtain meaningful information on the dependence of velocity distribution on physical parameters such as kinematic viscosity and system size. 

\section{Methods}
\subsection{Langevin Navier-Stokes equation}
The velocity distribution in plant cells along the vertical line is illustrated in Fig. \ref{fig-1}(b). In actual plant cells, a circular flow  is rotating on the cell surface. This flow is not written on the tube. 
Figure \ref{fig-1}(c) is the computational domain obtained by simplifying the flow field of protoplasmic streaming to a 2D flow field. $V_B$ denotes the boundary velocity corresponding to the circular flow on the cell surface. This two-dimensional (2D) flow field is called Couette flow if the flow is driven only by the boundary velocity. In this case, the Navier-Stokes equation is trivial because it has the exact solution. However, we assume that fluid particles thermally fluctuate, and this fluctuation is called Brownian motion. Thus,  the protoplasmic streaming is activated by both the boundary velocity and Brownian force from a fluid mechanical viewpoint. Due to the Brownian motion, the velocity distribution of $V_x$ or $V$ has two peaks at $V\!\to\!0$ and $V(\not=\!0)$, as shown in Fig. \ref{fig-1}(d). Since $h(V)$ is considered to be a probability distribution of $V$, these peaks indicate that many fluid particles are of $V\!\to\!0$ and $V\!=\!{\rm finite}$, which corresponds to $V_B$ in the case of Couette flow.

The random Brownian motion of fluid particles is naturally described by 
\begin{eqnarray}
\label{NS-eq-org}
\begin{split}
&\frac{\partial {\vec V}}{\partial t}=-\left ({\vec V}\cdot \nabla\right){\vec V}-{\rho}^{-1} {\it \nabla} p +\nu \Laplace {\vec V} + {\vec \eta},\\
&\nabla\cdot {\vec V}=0,
\end{split}
\end{eqnarray}
where ${\vec V}\!=\!(V_x,V_y)$ is the velocity of the fluid particle, and $p$ is the pressure (see also Ref. \cite{Ermak-McCammon-JCP1978} for Brownian dynamics of particles). 
The differential operators $\nabla$ and $\Laplace$ are defined by $\nabla\!=\!(\partial/\partial x, \partial/\partial y)$ and $\Laplace\!=\!\partial^2/\partial x^2\!+\! \partial^2/\partial y^2$, respectively. 
 The symbols $\rho$ and $\nu$ are density $(\rm{kgm^{-3}})$ and kinematic viscosity $(\rm{m^2s^{-1}})$, respectively. The final term $\vec{\eta}$ in the first of Eq. (\ref{NS-eq-org}) denotes Brownian force, given by Gaussian random numbers in the numerical simulations. The second equation is a condition, the divergence-less of velocity, for flows to be incompressible. We call these equations Langevin Navier-Stokes (LNS) equations, which will be numerically solved with the Marker and Cell (MAC) method on two different staggered lattices~\cite{McKee-etal-CandF2007}. Details of the staggered lattices are presented in the following subsection. 
 The LNS equation is numerically solved under the condition ${\partial {\vec V}}/{\partial t}\!=\!0$.

First of all, we should note that NS equation with the vortex $\omega(=\!(\nabla \times \vec{V})_z)$ and the stream function $\psi$ 
\begin{eqnarray}
\label{NS-eq-vortex}
\begin{split}
&\frac{\partial \omega}{\partial t}=-\left ({\vec V}\cdot \nabla\right)\omega +\nu \Laplace {\omega} + \nabla\times {\vec \eta},\\
&\omega=-{\it \Delta} \psi,
\end{split}
\end{eqnarray}
is obtained by applying $\nabla \times$ to the NS equation in Eq. (\ref{NS-eq-org}). Therefore, the vortex $\omega$ of velocity $\vec{V}$, which is a solution to the NS equation in Eq. (\ref{NS-eq-org}), satisfies the NS equation in Eq. (\ref{NS-eq-vortex}). However, the converse is not always true. Therefore, it is non-trivial whether the NS equation in Eq. (\ref{NS-eq-org}) has two-different peaks in the velocity distribution in Couette flow in parallel plates.

The variables velocity $\vec{V}$ and pressure $p$ in the LNS equation in Eq. (\ref{NS-eq-org}) are different from flow function $\psi$ and vorticity $\omega$ in the LNS equation simulated in Ref.~\cite{Egorov-etal-POF2020}, where the condition $\nabla\cdot {\vec V}\!=\!0$ is exactly satisfied. In contrast, this condition is not always satisfied in the time evolution of Eq. (\ref{NS-eq-org})  even though it is satisfied in the initial configuration.  The original MAC method is a simple technique to resolve this problem \cite{McKee-etal-CandF2007}, however,  $\nabla\cdot {\vec V}\!=\!0$ is not always satisfied, and hence, a simplified MAC (SMAC) method, which is well-known, is used in the simulations. In this technique,  $\nabla\cdot \vec {V}\!=\!0$ is successfully obtained in the convergent solutions corresponding to $\partial \vec{V}/\partial t\!=\!0$.   Here, we briefly introduce this technique. 

To discretize the time derivative in Eq. (\ref{NS-eq-org}) with the time step ${\it \Delta} t$, we have
\begin{eqnarray}
\label{NS-eq-time-step}
{\vec V}(t+{\it \Delta} t)= \vec{V}(t) +{\it \Delta} t \left[ \left (-{\vec V}\cdot \nabla\right){\vec V}(t)-{\rho}^{-1} {\it \nabla} p(t+{\it \Delta} t) +\nu \Laplace {\vec V}(t)\right] +\sqrt{2D{\it \Delta} t} \,\vec{g}(t),
\end{eqnarray}
where $\sqrt{2D{\it \Delta} t} \,\vec{g}\!=\!\vec{\eta}{\it \Delta} t$ (see Ref. \cite{Egorov-etal-POF2020} for more detailed information on this point). 
From this time evolution equation, we understand that $\nabla\cdot {\vec V}(t+{\it \Delta} t)\!=\!0$ is not always satisfied even if $\nabla\cdot {\vec V}(t)\!=\!0$ is satisfied because  the terms independent of ${\vec V}(t)$ in the right hand side are not always divergence-less. Moreover, the time evolution of $p(t)$ is not specified. For these reasons, we introduce a temporal velocity $\vec{V}^*(t)$ and rewrite Eq. (\ref{NS-eq-time-step}) as follows:
\begin{eqnarray}
\label{NS-eq-time-step-temporal-1}
&&{\vec V}^*(t)= \vec{V}(t) +{\it \Delta} t \left[ \left (-{\vec V}\cdot \nabla\right){\vec V}(t)-{\rho}^{-1} {\it \nabla} p(t) +\nu \Laplace {\vec V}(t)\right] +\sqrt{2D{\it \Delta} t} \,\vec{g}(t),\\
\label{NS-eq-time-step-temporal-2}
&&{\vec V}(t+{\it \Delta} t)={\vec V}^*(t)-{\it \Delta} t {\rho}^{-1} {\it \nabla} \left[ p(t+{\it \Delta} t)-p(t)\right].
\end{eqnarray}

By applying the operator $\nabla\cdot$ to Eq. (\ref{NS-eq-time-step-temporal-2}), 
we have
\begin{eqnarray}
\label{SMac-eq-0}
\nabla\cdot{\vec V}(t+{\it \Delta} t)=\nabla\cdot{\vec V}^*(t)-{\it \Delta} t {\rho}^{-1} {\it \Delta} \left[ p(t+{\it \Delta} t)-p(t)\right].
\end{eqnarray}
Then, assuming the condition $\nabla\cdot{\vec V}(t+{\it \Delta} t)\!=\!0$, 
we obtain the Possion's equation for $\phi(t) \!=\! p(t+{\it \Delta} t)\!-\!p(t)$ such that 
\begin{eqnarray}
\label{SMac-Poiss}
\Delta \phi(t)=\frac{\rho}{{\it \Delta} t}\nabla\cdot{\vec V}^*(t),\quad \phi(t) = p(t+{\it \Delta} t)-p(t).
\end{eqnarray}
Thus, combining Eq. (\ref{NS-eq-time-step-temporal-1}) for the time evolution of ${\vec V}^*(t)$ with the Possion's equation in Eq. (\ref{SMac-Poiss}) for  $\phi(t) \!=\! p(t+{\it \Delta} t)\!-\!p(t)$ and replacing $p(t)$ with $p(t)\!+\!\phi(t)$, we implicitly obtain the time evolution ${\vec V}(t+{\it \Delta} t)$ with the condition $\nabla\cdot{\vec V}(t+{\it \Delta} t)\!=\!0$. This technique to update  ${\vec V} (t)$ is slightly different from that of original MAC method, where ${\vec V} (t)$  is explicitly updated to ${\vec V}(t+{\it \Delta} t)$, and hence, $\nabla\cdot{\vec V}(t+{\it \Delta} t)\!=\!0$ is not always satisfied or slightly violated. This violation becomes larger for larger Brownian force strength $D$ and persists even in the convergent configurations in the original MAC method. 

To make clear the condition $\nabla\cdot{\vec V}\!=\!0$, we include it in the convergence condition.
The convergent criteria for the time step are 
\begin{eqnarray}
\label{convergent-time-step}
\begin{split}
&{\rm Max}\left[\left||\nabla\cdot\vec{V}_{ij}(t+{\it \Delta} t)|-|\nabla\cdot\vec{V}_{ij}(t)|\right|\right]<1\times 10^{-8},\\
&{\rm Max}\left[|\vec{V}_{ij}(t+{\it \Delta} t)-\vec{V}_{ij}(t)|\right]<1\times 10^{-8},\\
&{\rm Max}\left[|p_{ij}(t+{\it \Delta} t)-p_{ij}(t)|\right]<1\times 10^{-8},
\end{split}
\end{eqnarray}
and that of  iterations for the Poisson's equation is
\begin{eqnarray}
\label{convergent-Poisson}
{\rm Max}\left[|\phi_{ij}(n+1)-\phi_{ij}(n)|\right]<1\times 10^{-10},
\end{eqnarray}
where $n$ denotes the iteration step for the Poisson's equation in Eq. (\ref{SMac-Poiss}). The suffix $ij$ denotes a lattice site ranging $1\!\leq\! i\!\leq\! n_X, 1\!\leq\! j\!\leq\! n_Y$ (the lattice structure is presented in the following subsection). 
Acceleration coefficient $A\!=\!1.87$ is assumed for the iteration of the Poisson's equation. The total number of convergent configurations  is approximately $2.5\times 10^4$, which is used for the mean value calculation of all physical quantities.

We should note that the first condition in Eq. (\ref{convergent-time-step}) is immediately satisfied in the early stage of iterations, and the final value of ${\rm Max}\left[\left||\nabla\cdot\vec{V}_{ij}(t+{\it \Delta} t)|-|\nabla\cdot\vec{V}_{ij}(t)|\right|\right]$ for the convergent configuration is $1\times 10^{-14}$ or less, 
and therefore, this convergence condition is actually unnecessary. One more point to note is that only convergent solution satisfies $\nabla\cdot{\vec V}\!=\!0$. In this sense, the obtained  numerical solution of LNS equation in Eq. (\ref{NS-eq-org}) is a steady state solution characterized by ${\partial \vec{V}}/{\partial t}\!=\!0$ as mentioned above.

The LNS equation in Eq. (\ref{NS-eq-org}) is a stochastic equation, and therefore, the mean value of physical quantity $Q$ is obtained by 
\begin{eqnarray}
\label{mean-value}
 Q=(1/N_{\rm sa})\sum_{n=1}^{N_{\rm sa}} Q(n),
\end{eqnarray}
where $Q(n)$ is a quantity calculated from the $n$-th convergent solution $\{ \vec{V}_{ij}(n)\}$.  The total number $N_{\rm sa}$ of samples is $N_{\rm sa}\!=\!1$ for $D\!=\!0$, and $N_{\rm sa}\simeq 2.5\times10^4$ for $D\not=0$.

\subsection{Staggered lattice}
\begin{figure}[!h]
\centering{}\includegraphics[width=12cm]{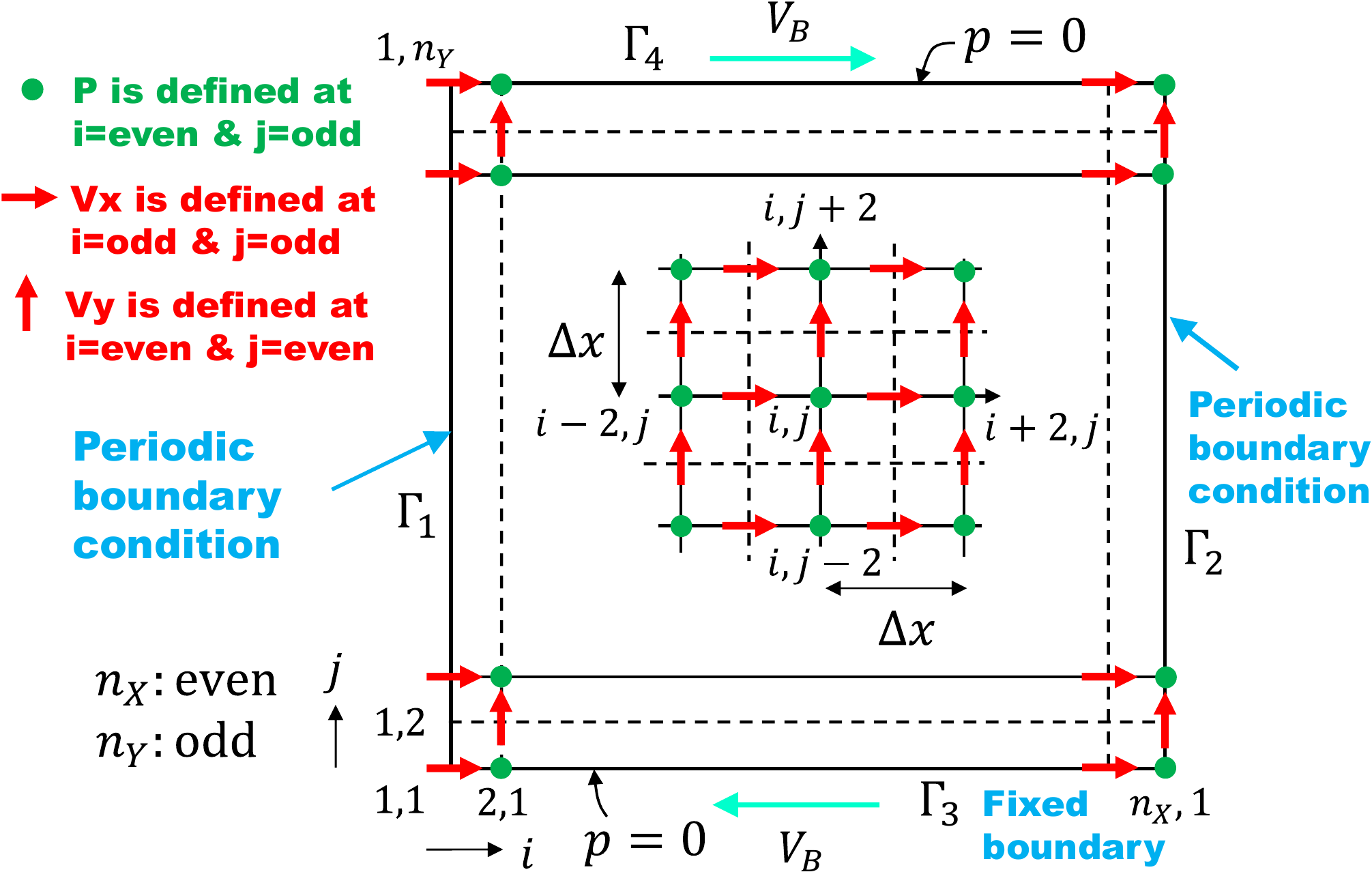}
\caption{
Regular square and staggered lattice  of size $(n_X,n_Y)$ ($n_X\!={\rm even}$, $n_Y\!={\rm odd}$), where pressure $p$ ($\textcolor{green}{\bullet}$) and velocity $V_x$ ($\textcolor{red}{\rightarrow}$), $V_y$ ($\textcolor{red}{\uparrow}$) are defined alternatively at different points, and the lattice spacing ${\it \Delta} x$ is twice larger than the standard lattice. Periodic boundary condition is assumed on the left ($\Gamma_1$)  and right ($\Gamma_2$)  boundaries at $i\!=\!1$ and $i\!=\!n_X$ such that $(n_X\!+\!1,j)$ is identified with $(1,j)$. Dirichlet boundary condition is assumed for $p$ on the upper ($\Gamma_4$) and lower ($\Gamma_3$)  boundaries, where $\vec{V}$ is fixed to $\vec{V}\!=\!(\pm V_B,0)$, and $n_X\!=\!100$ and $n_Y\!=\!101$ in the simulations.
}
\label{fig-2}
\end{figure}
We introduce a staggered lattice  \cite{McKee-etal-CandF2007} for space discretization of  Eq. (\ref{NS-eq-time-step}) (Figs. \ref{fig-2}). The lattice size is given by the total number of vertices $N\!=\!n_X\!\times\! n_Y$, where $n_X\!=\!100$ and $n_Y\!=\!101$ are assumed in the simulations. We should note that $n_X$ is effectively half of that assumed in Ref.~\cite{Egorov-etal-POF2020}, because the sites where $\vec{V}$ is defined are different from those where $p$ is defined, as shown in Fig. \ref{fig-2}. The lattice spacing ${\it \Delta} x$ is given by the smallest distance between the sites in which the same variable is defined, and therefore, the side length $[(n_X\!-\!2)/2]\!\times\!{\it \Delta}x$ of the staggered lattice is the same as that of the standard lattice in Ref.~\cite{Egorov-etal-POF2020} for the same ${\it \Delta}x$. We should note that only half of the lattice points are used for the variables on the staggered lattice.

Periodic boundary condition (PBC) is assumed on the boundaries $\Gamma_1$ and $\Gamma_2$. This  PBC implies that $\vec{V}(n_x\!+\!1,j)\!=\!\vec{V}(1,j)$ and $p(n_x\!+\!1,j)\!=\!p(1,j)$ for all $j$. For the pressure $p$, Dirichlet boundary condition $p\!=\!0$ is assumed on $\Gamma_3$ and $\Gamma_4$ on both lattices.  $V_y$ is fixed to $V_y\!=\!0$ at $(i,1),(i=1,\cdots,n_X)$ and $(i,n_Y-1),(i=1,\cdots,n_X)$, where $n_Y\!=\!n_X\!+\!1$.

\subsection{Dimensional analysis \label{dim-analysis}}
Due to the non-standard dependence of the LNS equation in Eq. (\ref{NS-eq-time-step}) on ${\it \Delta} t$,  special attention should be paid to the numerical solution not only on ${\it \Delta} t$ but also on  ${\it \Delta} x$. As discussed in Ref.~\cite{Egorov-etal-POF2020}, the solution should be independent of ${\it \Delta} t$, ${\it \Delta} x$ and simulation units, which are always different from the real units for length $({\rm m})$, time $({\rm s})$, and weight $({\rm kg})$, where the final one  $({\rm kg})$ is necessary for the LNS equation in Eq. (\ref{NS-eq-time-step}), because it includes $\rho ({\rm kgm^{-3}})$, which is not included in LNS equation in Ref.~\cite{Egorov-etal-POF2020}. 

The flow field changes with changing real physical parameters; diameter $d({\rm m})$ of plant cell, boundary velocity $V_B ({\rm ms^{-1}})$,  density $\rho ({\rm kgm^{-3}})$,  kinematic viscosity $\nu ({\rm m^2s^{-1}})$, relaxation time $\tau ({\rm s})$, and strength $D ({\rm m^2s^{-3}})$ of Brownian force. The time step  ${\it \Delta} t$  and lattice spacing ${\it \Delta} x$  are connected to $d$ and $\tau$ by
 \begin{eqnarray}
\label{dt-dx}
{\it \Delta} x=\frac{d}{n_Y}, \quad {\it \Delta} t=\frac{\tau}{n_T},
\end{eqnarray}
where $n_Y$ is the total number of lattice points in the vertical edge (Fig. \ref{fig-2}), and the total number of time iterations $n_T$ is also introduced to define  ${\it \Delta} t$ symmetrically with ${\it \Delta} x$. 

The simulation units are defined by using positive numbers $\alpha$, $\beta$, and $\lambda$ such that
\begin{eqnarray}
\label{sim-units}
\alpha {\rm m}, \quad \beta {\rm s}, \quad \lambda {\rm kg}.
\end{eqnarray}
Including the dependence of time step and lattice spacing, which can be changed by positive numbers $\gamma$ and $\delta$, we have a scale transformation  
\begin{eqnarray}
\label{scale-transf}
 {\rm m}, {\rm s}, {\rm kg}, n_X, n_T \to \alpha {\rm m}, \beta {\rm s},  \lambda {\rm kg}, \gamma n_X, \delta n_T,
\end{eqnarray}
where the final two are equivalent to ${\it \Delta} x \to \gamma^{-1}{\it \Delta} x$ and ${\it \Delta} t \to\delta^{-1}{\it \Delta} t$.  
Thus, it is reasonable to assume that the convergent solution of  Eq. (\ref{NS-eq-time-step})  is independent of the scale transformation in Eq. (\ref{scale-transf}). 

To see this scale invariance in more detail, we apply the transformation to Eq. (\ref{NS-eq-time-step}) , and find that Eq. (\ref{NS-eq-time-step}) is invariant if $\nu$ and $D$ scale according to 
\begin{eqnarray}
\label{parameter-scale-1}
\nu \to \gamma^{-1}\nu, \quad D \to \gamma^2\delta^{-1}D, \quad(\gamma=1),
\end{eqnarray}
where $\gamma\!=\!1$ is assumed. 
The detailed information on this part, including the reason for the assumption $\gamma\!=\!1$, is given in Appendix A. 

The basic considerations on the unit changes in Eq. (\ref{sim-units})  can be slightly extended. For this purpose, we use notions of a set of changeable physical parameters $E$ and a set of changeable simulation parameters $S$, which are respectively given by
\begin{eqnarray}
\label{parameter-sets}
E=(\rho_e, \nu_e,V_{Be},d_e),\quad  S=(\rho_0, \nu_0,V_{B0}, D_0, {\it \Delta}x_0, {\it \Delta}t_0).
\end{eqnarray}
The density $\rho_e$ is included in $E$, though $\rho_e$ is unchangeable and almost exactly the same as that of water and independent of plants and other conditions. Note that $\rho_0$ is changeable in simulations, and it should be included  in $S$. Let us denote experimentally observed data by ${\rm Exp}(E)$ corresponding to the parameter set $E$.

\begin{figure}[!h]
\centering{}\includegraphics[width=10cm]{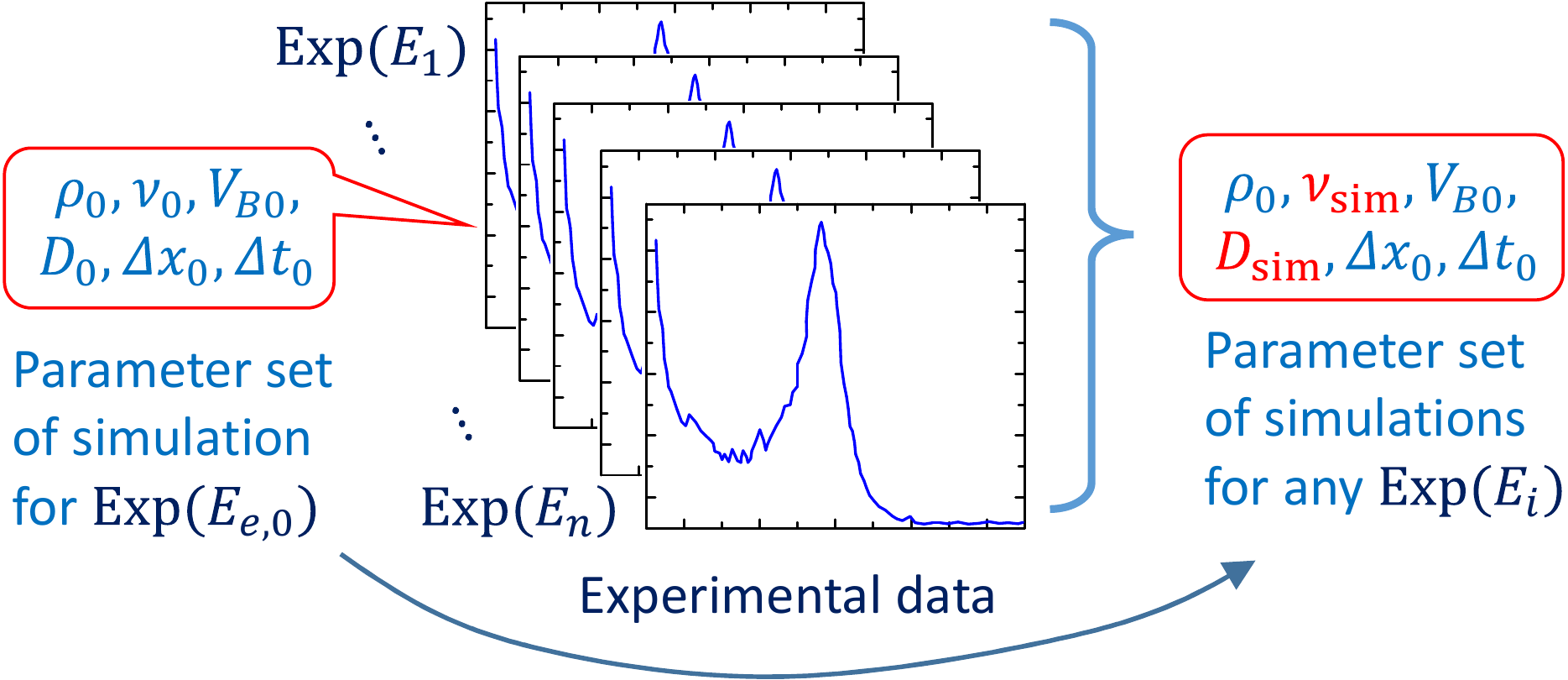}
\caption{An illustration of statement (B). A set of parameters $S\!=\!(\rho_0, \nu_0,V_{B0}, D_0, {\it \Delta}x_0, {\it \Delta}t_0)$ of a simulation for  experimental data of ${\rm Exp}(E_{e,0})$. By using this set of parameters, any experimental data corresponding to $E_i(=\!E_{e})$ can be simulated by replacing   $\nu_0$ and $D_0$ with suitable $\nu_{\rm sim}$ and $D_{\rm sim}$ in $S$.
}
\label{fig-3}
\end{figure}

The first statement describing the invariance of solution under the scale transformation in Eq. (\ref{scale-transf}) is
\begin{enumerate}
\item[(A)]The convergent solution $(\vec{V},p)$ of Eq. (\ref{NS-eq-time-step}) for a parameter set $S$ in Eq. (\ref{parameter-sets}) remains unchanged if and only if there exists a set of positive parameters $(\alpha,\beta, \lambda, \gamma,\delta)$ such that $S$ transforms to 
\begin{eqnarray}
\label{parameter-scale-2}
S\to\left(\frac{\alpha^3}{ \lambda}\rho_0, \frac{\beta}{\alpha^{2}\gamma}\nu_0,\frac{\beta}{\alpha} V_{B0}, \frac{\beta^3\gamma^2}{\alpha^{2}\delta}D_0, \frac{1}{\alpha\gamma}{\it \Delta}x_0,\frac{1}{\beta\delta} {\it \Delta}t_0\right), \quad(\gamma=1)
\end{eqnarray}
\end{enumerate}
This statement (A) is the same as that in Ref.~\cite{Egorov-etal-POF2020} except the fact that $S$ includes $\rho$ and the extra condition $\gamma\!=\!1$ in this paper. The second statement is as follows:
\begin{enumerate}
\item[(B)]Let $E_{e,0}=(\nu_{e,0},V_{Be,0},d_{e,0})$, $S\!=\!(\rho_0, \nu_0,V_{B0}, D_0, {\it \Delta}x_0, {\it \Delta}t_0)$. If ${\rm Exp}(E_{e,0})$ is successfully simulated with $S$, then for any ${\rm Exp}(E_{e})$, there uniquely exists $\nu_{\rm sim}$ and $D_{\rm sim}$ such that ${\rm Exp}(E_{e})$ can be simulated with $(\rho_0, \nu_{\rm sim},V_{B0}, D_{\rm sim}, {\it \Delta}x_0, {\it \Delta}t_0)$.
\end{enumerate}
This statement is illustrated in Fig. \ref{fig-3}. The statement (B) is slightly different from that in Ref.~\cite{Egorov-etal-POF2020}, because $\nu_0$ in $S$ in Eq. (\ref{parameter-scale-2}) should be replaced by $\nu_{\rm sim}$. This difference comes from the fact that the LNS equation in Eq. (\ref{NS-eq-time-step}) is not the same as the LNS equation in Ref.~\cite{Egorov-etal-POF2020}, as mentioned above. However, under the condition $\gamma\!=\!1$, the transformation rule for $\nu_0$ is $\nu\to \frac{\beta}{\alpha^{2}}\nu_0$ implying that $\nu_{\rm sim}$ can be replaced by $\nu_0$, and therefore, the statement (B) except for $\rho$ is the same as in  Ref.~\cite{Egorov-etal-POF2020}. In other words, results obtained by the statement (B) can be obtained by the statement (B) in  Ref.~\cite{Egorov-etal-POF2020} under the condition $\gamma\!=\!1$. In this sense, the range of invariance of the scale transformation is limited or narrow in the LNS equation in this paper though the statement (B) in this paper is still interesting. 


\section{Results}
\subsection{Physical and simulation parameters}
The problem is what type of protoplasmic streaming information can be extracted from LNS simulations in Eq. (\ref{NS-eq-org}). To answer this question, we can use statement (B). The assumption part of (B) is that ${\rm Exp}(E_{e,0})$ is successfully simulated with $S$, and hence, we assume physical values for $E_{e,0}$ in Table \ref{table-1}. These are from  Refs.~\cite{KamiyaKuroda-BMT1956,KamiyaKuroda-BMT1973} and are the same as assumed in Ref.~\cite{Egorov-etal-POF2020}. The numerical results will be presented in the following subsection to confirm that the assumption part is correct.
\begin{table}[!ht]
\centering
\caption{
The assumed physical parameters are the same as those in Ref.~\cite{Egorov-etal-POF2020} except the density $\rho_{e,0}$. These are from Refs.~\cite{KamiyaKuroda-BMT1956,KamiyaKuroda-BMT1973}. The density $\rho_{e,0}$ is the same as that of water, and the kinematic viscosity $\nu_{e,0}$  is approximately 100 times larger than that of water. 
}
\begin{tabular}{|c|c|c|c|c|c|c|c|}
\hline
\multicolumn{4}{|c|}{\bf Physical parameters} \\  \hline
 $\rho_{e,0}({\rm {kg}/m^3})$ & $\nu_{e,0}({\rm {m^2}/{s}})$ & $d_{e,0} ({\rm m})$ & $V_{Be,0} ({\rm m}/s)$ \\ \hline
1000  & $1\times10^{-4} $ & $5\times10^{-5}$ & $5\times10^{-4}$ \\ \hline
\end{tabular}
\label{table-1}
\end{table}
The simulation parameters $S$ are obtained by fixing the simulation units, which are defined by a set of positive numbers $\alpha_0, \beta_0, \gamma_0, \lambda_0, \delta_0$ in Table \ref{table-2}. 
\begin{table}[!ht]
\centering
\caption{
These parameters define the simulation units for simulating ${\rm Exp}(E_{e,0})$, where $E_{e,0}=(\rho_{e,0},\nu_{e,0},d_{e,0},V_{Be,0})$ in Table \ref{table-1}. 
 }
\begin{tabular}{|c|c|c|c|c|c|c|c|}
\hline
 \multicolumn{5}{|c|}{\bf Positive numbers for the simulation unit}\\ \hline
$\alpha_0$ & $\beta_0$ & $\lambda_0$ & $\gamma_0$ & $\delta_0$  \\ \hline
$1\times10^{-6}$  & $0.1$ & $1\times10^{-12}$ & $1$ & 1  \\ \hline
\end{tabular}
\label{table-2}
\end{table}
In Table \ref{table-3}, the assumed simulation parameters $S$ for simulating ${\rm Exp}(E_{e,0})$ are shown. 
\begin{table}[!ht]
\centering
\caption{
The assumed  simulation parameters $S$, which are obtained by $\alpha_0, \beta_0, \gamma_0, \lambda_0,\delta_0$ in Table \ref{table-2}.  ${\it \Delta}x_{0}\!=\!10 (\alpha_0 {\rm m})$ is twice larger than that assumed in Ref.~\cite{Egorov-etal-POF2020}, which implies $n_X(=\!d_0/{\it \Delta}x_{0})\!=\!50$ on the standard lattice.  Note that $n_X$ is on the staggered lattice in Fig. \ref{fig-2} is $n_X\!=\!100$ because the lattice spacing ${\it \Delta}x_{0}$ is twice larger than that on the standard lattice.}
\begin{tabular}{|c|c|c|c|c|c|c|c|}
\hline
\multicolumn{6}{|c|}{\bf Assumed simulation parameters} \\ \hline
$\rho_{0} (\frac{\lambda_0{\rm {kg}}}{\alpha_0^3m^3})$ & $\nu_{0}({\rm \frac{\alpha_0^2m^2}{\beta_0s}})$ &   $V_{B0}({\rm \frac{\alpha_0 m}{\beta_0s}})$ & $d_{0}({\rm \alpha_0 m})$  & ${\it \Delta}x_{0} ({\rm \alpha_0 m})$  & ${\it \Delta}t_{0} ({\rm \beta_0s})$ \\ \hline
$1\times 10^{-3}$ & $1\times10^{7} $ & 5 & 500 & 5 & $5\times 10^{-7}$ \\ \hline 
\end{tabular}
\label{table-3}
\end{table}

\subsection{Velocity distributions}
As we have discussed in the first of the preceding subsection, the remaining task to be performed  is to check the assumption part of statement (B). The assumed simulation parameters are listed in Table \ref{table-3}. The parameter ${\it \Delta}x_0$ given by Eq. (\ref{dx-dt-scale-trans}) such that ${\it \Delta}x_0=\frac{d_{e,0}}{\gamma n_X}\frac{\nu_0}{\nu_{e,0}}\frac{V_{e,0}}{V_0} \;(\alpha {\rm m})$, where $E_{e}$ is replaced by $E_{e,0}$. The time step ${\it \Delta}t_0$ is logically obtained by the second of Eq. (\ref{dx-dt-scale-trans}), however, $\tau_{e,0}$ is experimentally unknown, and therefore, we assume a suitable number for  ${\it \Delta}t_0$. 

\begin{figure}[!h]
\centering{}\includegraphics[width=12.0cm]{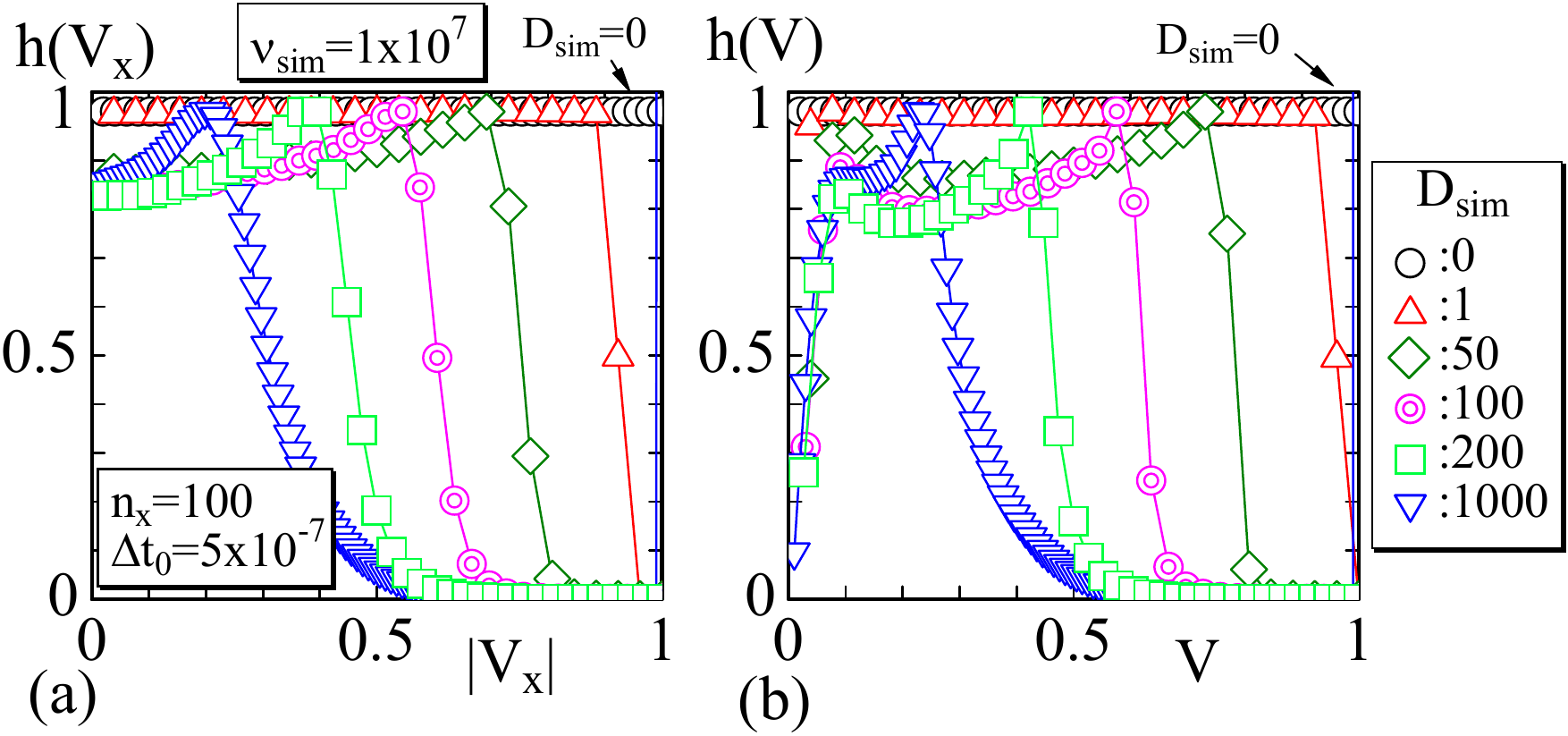}
\caption{{ Normalized velocity distribution on Lat 1.}  (a) Normalized velocity distribution $h(V_x)$ vs. $|V_x|$ and (b) $h(V)$ vs. $V$ obtained on Lat 1. 
Kinematic viscosity is fixed to $\nu_{\rm sim}\!=\!1\!\times\! 10^7$, and $D_{\rm sim}$ is varied in the range $0\!\leq\! D_{\rm sim}\!\leq\! 1000$. }
\label{fig-4}
\end{figure}

Figures \ref{fig-4}(a) and (b) show distributions or normalized  histograms of velocities $h(V_x)$ vs. $|V_x|$ and $h(V)$ vs. $V(=\!|\vec{V}|)$. The kinematic viscosity is fixed to $\nu_{\rm sim}\!=\!\nu_0$, while random force strength is varied in the range $0\!\leq\!D_{\rm sim}\!\leq\!1000$. Both $h(V_x)$ and $h(V)$ have two peaks at zero and finite velocities, and the position of peaks at finite velocities moves left as $D_{\rm sim}$ increases. These features are the same as those observed in Ref.~\cite{Egorov-etal-POF2020}.  A peak at finite $|V|$ appears in $h(V)$ for the region larger than $D_{\rm sim}\!\simeq\!10$ at least, which is not shown in Fig. \ref{fig-4}, and these peak values $h(V)$ are higher than those at $|V|\!\to\!0$ in contrast to the case of the LNS equation in Ref~\cite{Egorov-etal-POF2020}.  Besides this problem of peak heights, two different peaks are clearly observed in  $h(V)$ at $D_{\rm sim}\!=\!50,100,200,1000$. Thus, we find that the assumption part of statement (B) is correct. 

\begin{figure}[!h]
\centering{}\includegraphics[width=10.0cm]{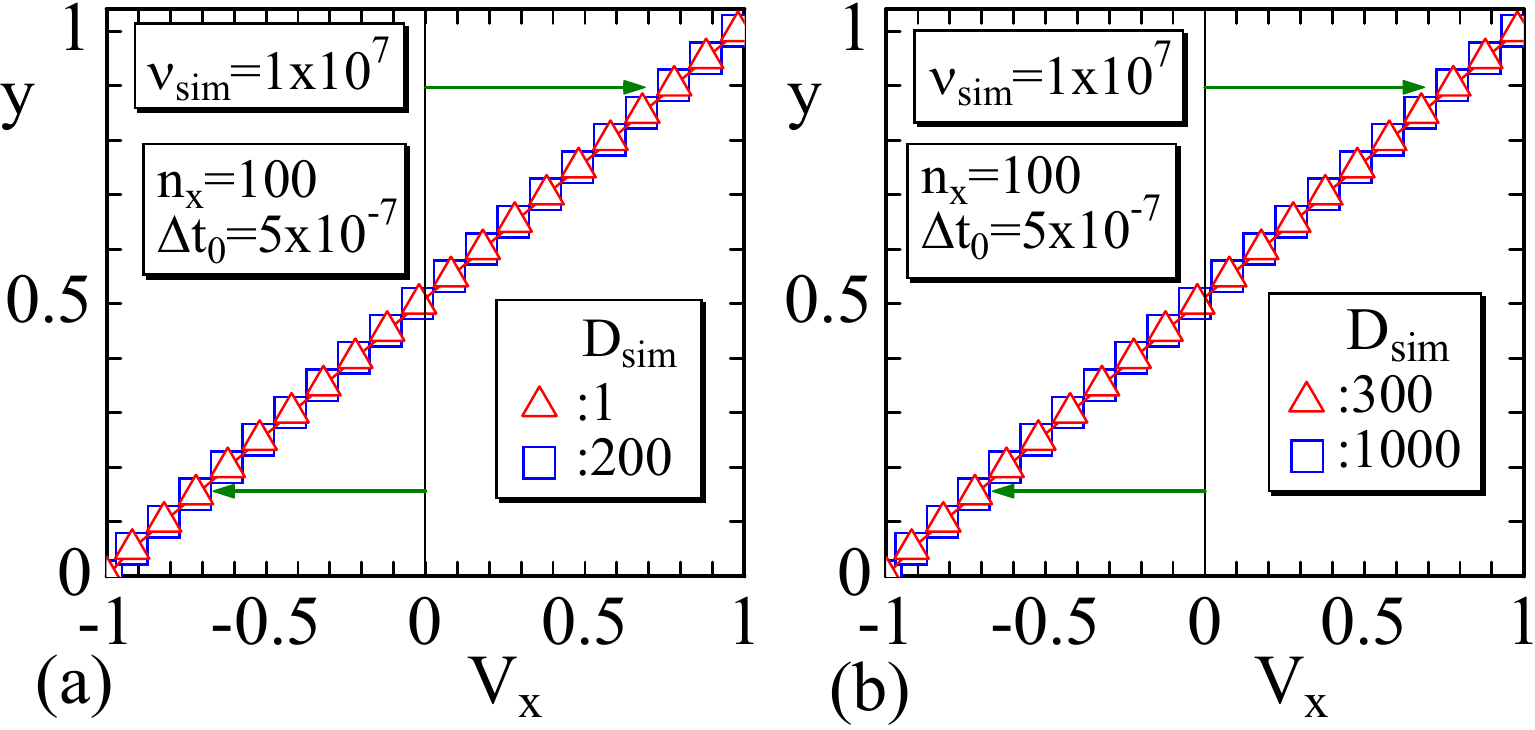}
\caption{Position dependence of velocity. Velocity $V_x$ vs. $y$ obtained for (a) $D_{\rm sim}\!=\!1,200$ and (b) $D_{\rm sim}\!=\!300,1000$. These are exactly the same as those in Ref.~\cite{Egorov-etal-POF2020}}
\label{fig-5}
\end{figure}

The position dependence of $V_x$ on $y$ is plotted in Figs. \ref{fig-5}(a) and (b). The results are almost the same as the exact solution $V_x\propto y$ for the case of zero Brownian force $D_{\rm sim}\!=\!0$. The fact that $V_x\propto y$ is exactly the same as in the simulation results of the LNS equation for vorticity and stream function in Ref.~\cite{Egorov-etal-POF2020}.

%

\subsection{Dependence on time discretization.} 
\begin{figure}[!h]
\centering{}\includegraphics[width=10.0cm]{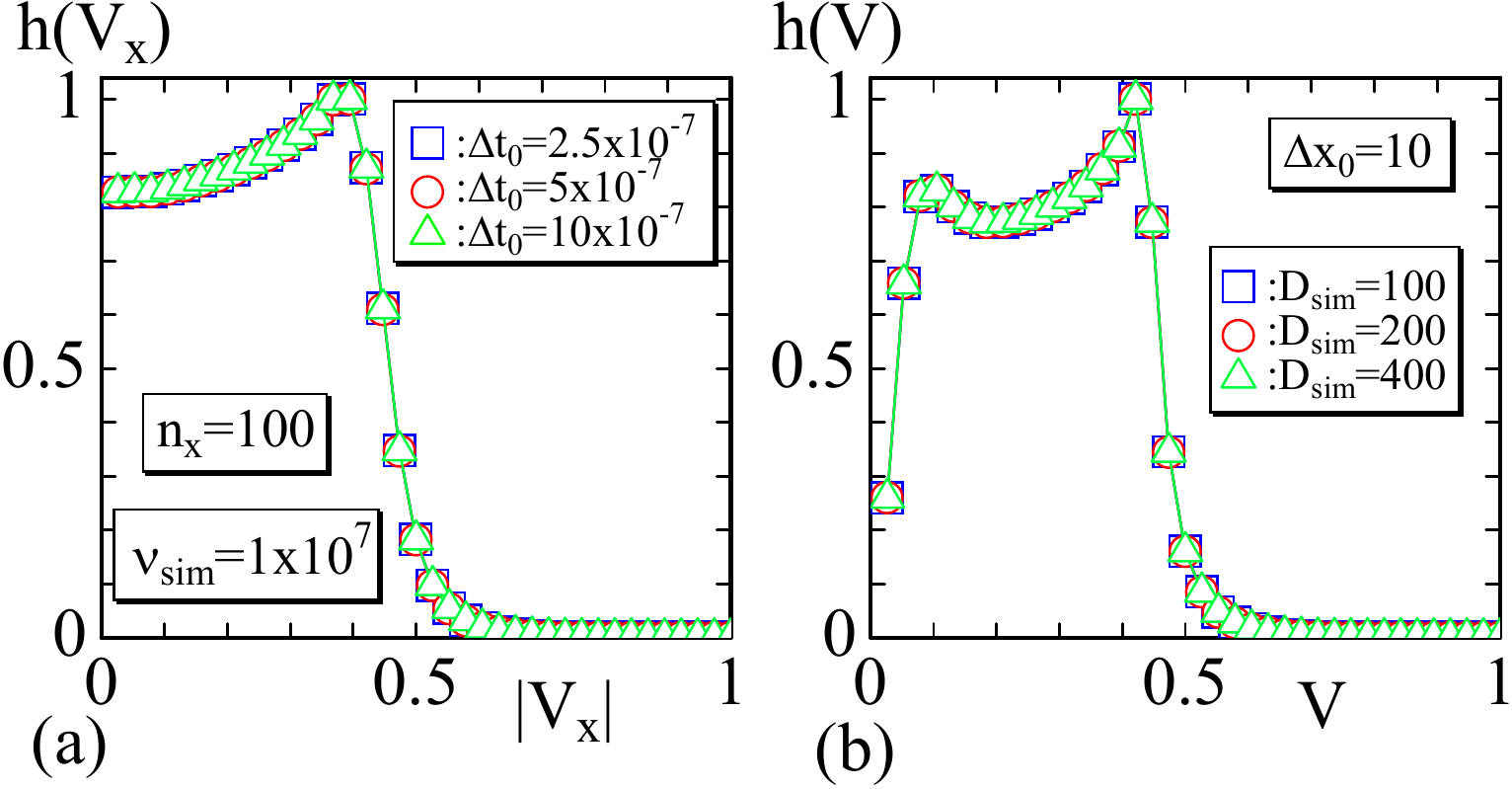}
\caption{
 Dependence of (a) $h(V_x)$ and (b) $h(V)$ on time discretization ${\it \Delta}t_0$, which is fixed to $2.5\!\times\! 10^{-7}$,   $5\!\times\! 10^{-7}$ and $10\!\times\! 10^{-7}$,   and correspondingly $D_{\rm sim}$ is fixed to  $D_{\rm sim}\!=\!100, 200$ and 400, respectively. The other parameters are the same as those assumed for  Fig. \ref{fig-2}.
Normalized velocity distributions (a) $h(V_x)$ vs. $|V_x|$ and  (b)  $h(V)$ vs. $V$. }
\label{fig-6}
\end{figure}
In the remaining part of this section, we check that our simulations are correctly performed. First, we check the time dependence corresponding to the parameter $\delta$ in ${\it \Delta}t\!\to\! \delta^{-1}{\it \Delta}t$ in Eq. (\ref{scale-transf}). To see the dependence of the results on $\delta_0$, we fix $\delta_0\!=\!1.5$ and $\delta_0\!=\!2$, respectively, and all other parameters $\alpha_0, \beta_0, \gamma_0, \lambda_0$ remain unchanged. This implies that the physical parameters remain unchanged  from the statement (A), and only time step changes by $\delta_0$. 
Accordingly, $D_0$ and ${\it \Delta}t_{0}$ are changed to those listed in the upper part of Table \ref{table-5}. In Table \ref{table-5}, $\nu_{\rm sim}$ and ${\it \Delta}x_0$ are also shown, where ${\it \Delta}t_0$ varies with varying $\delta_0$. From the results $h(V_x)$ and $h(V)$ plotted in Figs. \ref{fig-6}(a), (b),  the invariance under  ${\it \Delta}t\!\to\! \delta^{-1}{\it \Delta}t$ is confirmed.  

Checks for the lattice size dependence are unnecessary  because the lattice size change due to $\gamma$ is excluded from the scale transformation in Eq. (\ref{scale-transf}).

\begin{table}[!ht]
\centering
\caption{
 The assumed  simulation parameters $\nu_{\rm sim}$, $D_{\rm sim}$, ${\it \Delta}x_0$, and ${\it \Delta}t_0$ to check the time  discretization dependence.
 }
 
\begin{tabular}{|c|c|c|c|c|c|c|}
\hline
\multicolumn{7}{|c|}{\bf Parameters for time discretization} \\ 
\hline
$\delta_{0}$ & $\nu_{\rm sim}({\rm \frac{\alpha_0^2m^2}{\beta_0s}})$& $V_{B0}({\rm \frac{\alpha_0m}{\beta_0s}})$  & $D_{\rm sim}({\rm \frac{\alpha_0^2m^2}{\beta_0^3s^3}})$  & ${\it \Delta}x_{0} ({\rm \alpha_0 m})$  & ${\it \Delta}t_{0}({\rm \beta_0s})$ &$n_X$ \\ \hline
2&  $1\times10^{7} $  &5& $100$ & $10$ & $2.5\times 10^{-7}$ &100\\ \hline
1 &  $1\times10^{7} $ &5 & $200$ & $10$ & $5\times 10^{-7}$ &100\\ \hline
0.5 &  $1\times10^{7}$ &5 & $400$ & $10$ & $10\times 10^{-7}$ &100 \\ \hline
\end{tabular}
\label{table-5}

\end{table}

\begin{figure}[!h]
\centering{}\includegraphics[width=10.0cm]{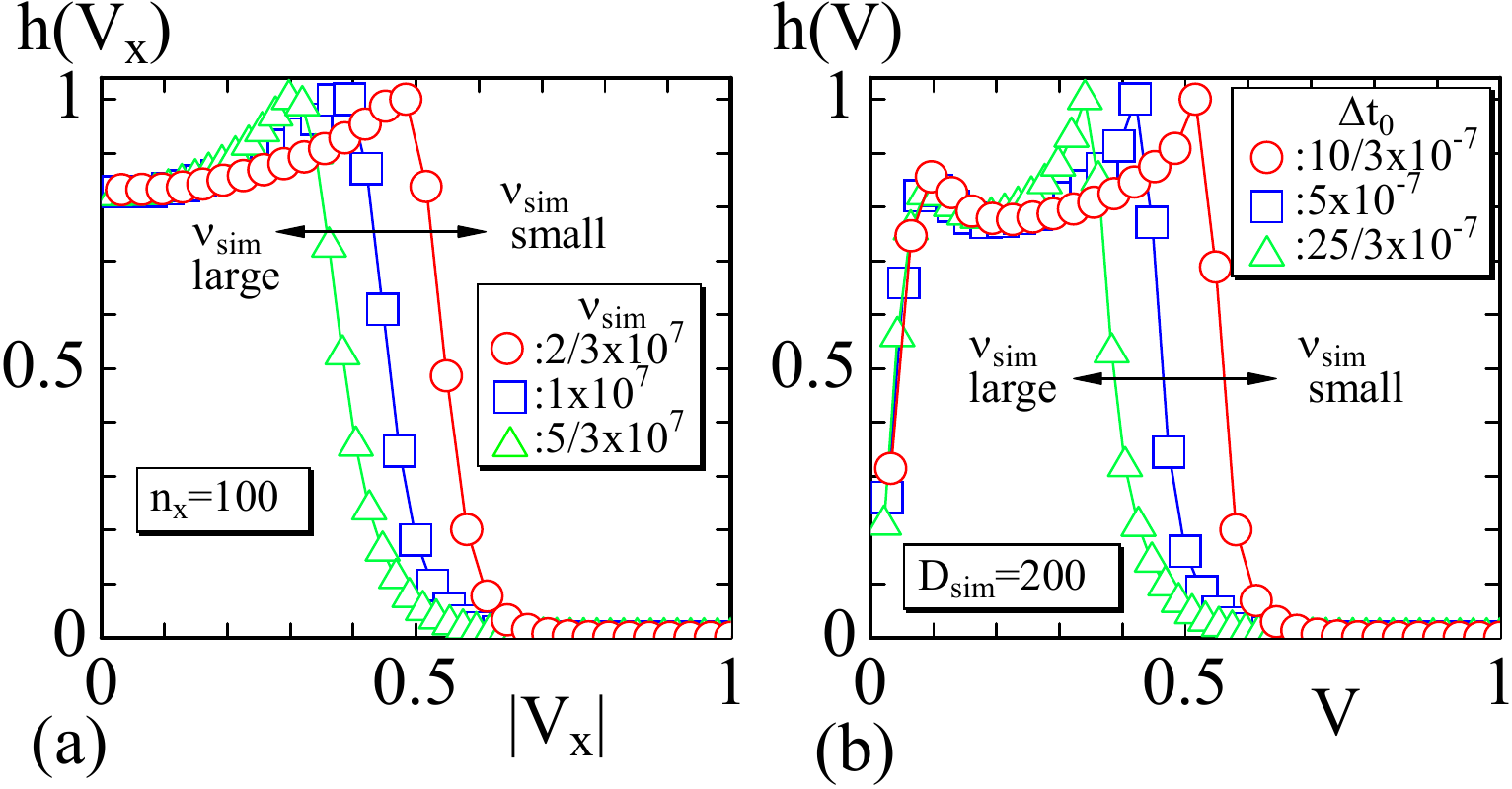}
\caption{Dependence of (a) $h(V_x)$ and (b) $h(V)$ on kinematic viscosity $\nu_{\rm sim}$,  
which  is varied to $2/3\!\times\!10^7$, $1\times\!10^7$, and $5/3\!\times\!10^7$, while $D_{\rm sim}$ is fixed to $D_{\rm sim}\!=\! 200$. Accordingly, the lattice spacing ${\it \Delta}x_0$ is also varied to $10/3$, $5$ and $25/3$. }
\label{fig-7}
\end{figure}

Finally, in this subsection, we show the dependence of velocity distribution on $\nu_{\rm sim}$. Note that the parameter $\nu_{\rm sim}$ is not included in the LNS equation in Ref.~\cite{Egorov-etal-POF2020}. Therefore, there is no result on the dependence of the peaks in $h(V_x)$ and $h(V)$ on  $\nu_{\rm sim}$  in Ref.~\cite{Egorov-etal-POF2020}. However, this dependence is non-trivial in the model of this paper, and hence, we study this problem. Figures \ref{fig-7}(a) and (b) show $h(V_x)$ and $h(V)$, where $\nu_{\rm sim}$ is varied. Since ${\it \Delta}x$ and ${\it \Delta}t$ depend on $\nu_0$, these values also change according to Eq. (\ref{dx-dt-scale-trans}) in Appendix \ref{APP-B}. We find that the peak position moves left (right) with increasing (decreasing) $\nu_{\rm sim}$. 

%

\subsection{Snapshots of velocity and pressure.} 
\begin{figure}[!h]
\centering{}\includegraphics[width=10.0cm]{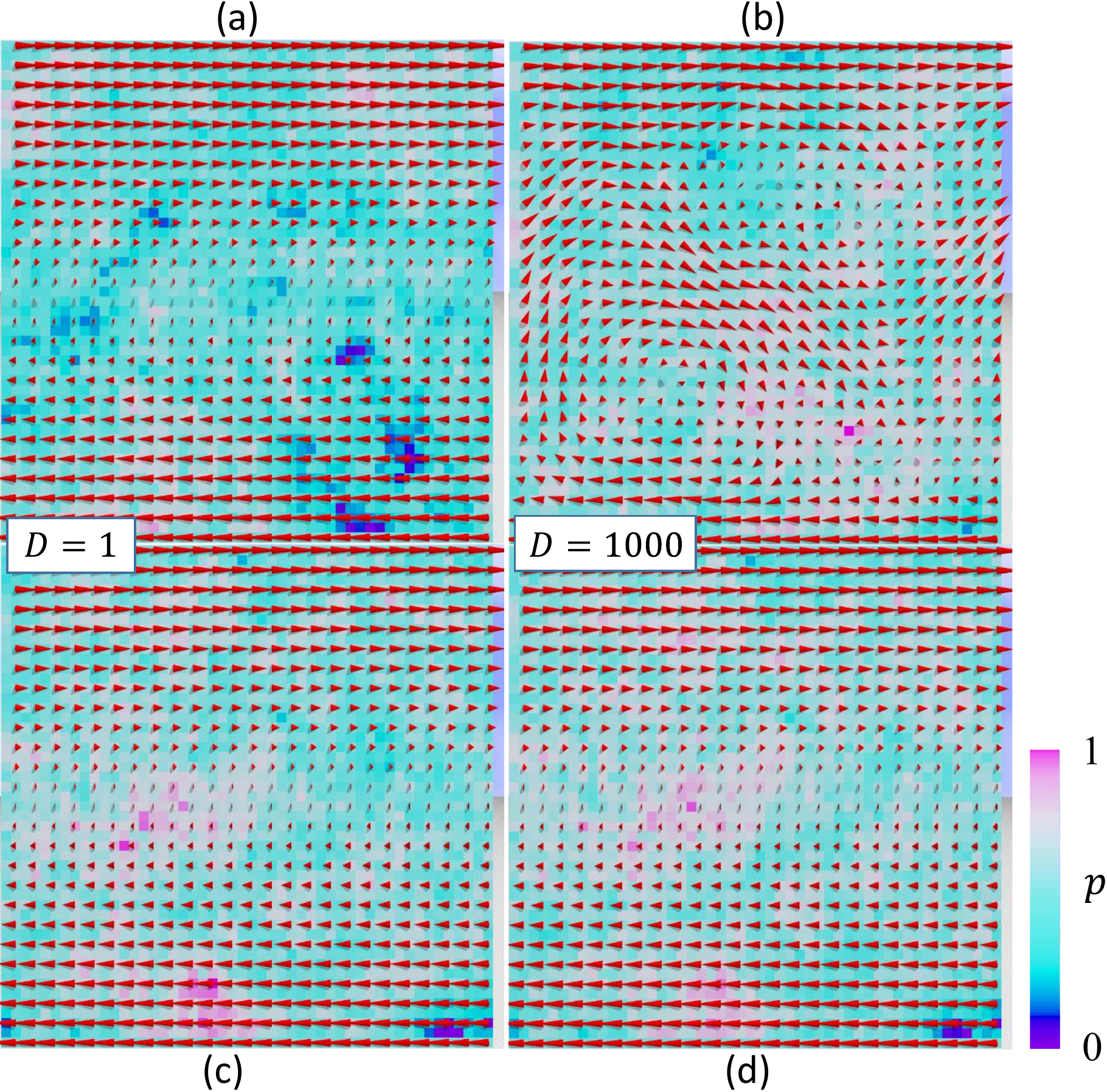}
\caption{Snapshots of velocity $\vec{V}$  and pressure $p$ of convergent configuration for (a) $D_{\rm sim}\!=\!1$ and (b) $D_{\rm sim}\!=\!1000$ shown in Fig. \ref{fig-4}. The velocity, shown at every other position, is represented by cones, and the pressure is normalized and visualized by blue-pink colors. The mean values $\vec{V}$ and $p$ of $2.5\!\times\!10^4$ convergent configurations are shown in (c) for $D_{\rm sim}\!=\!1$ and (d) for $D_{\rm sim}\!=\!1000$. The velocities $\vec{V}$ in (a),(c) and (d) are close the exact solution $(V_x,V_y)\!\sim\!(y,0)$, while $\vec{V}$ in (b) is not.  }
\label{fig-8}
\end{figure}

Snapshots of velocity $\vec{V}\!=\!(V_x,V_y)$ and pressure $p$ are shown in Figs. \ref{fig-8}(a) and (b). The pressure $p$ ranging $p_{\rm min}\!\leq\!p\!\leq\!p_{\rm max}$ is normalized such that $0\!\leq\! p\!\leq\! 1$, for visualization. This normalization is defined by $(p+|p|_{\rm max})/2|p|_{\rm max}$, where $|p|_{\rm max}\!=\!{\rm Max}[p_{\rm max},|p_{\rm min}|]$. Note that $p_{\rm min}\!<\!0$ and $p_{\rm max}\!>\!0$, and  $|p_{\rm min}|$ is not always equal to $p_{\rm max}$. Each of  $\vec{V}$ and $p$ is one of the convergent configurations corresponding to the data plotted in Fig. \ref{fig-4}.  The velocity configuration  in Fig.  \ref{fig-8}(a) for $D_{\rm sim}\!=\!1$ is close to the exact solution $(V_x,V_y)\!\sim\!(y,0)$ at  $D_{\rm sim}\!=\!0$, and no vortex can be seen. In contrast, the velocity in Fig.  \ref{fig-8}(b) for $D_{\rm sim}\!=\!1000$ considerably deviates from the exact solution, and a vortex-like configuration appears. Such a deviation is due to the Brownian motion and is considered the origin of the shape of $h(V_x)$ and $h(V)$ shown in Figs. \ref{fig-4}(a), (b). Hence, those vortex-like configurations are expected to play an important role in the mixing or transporting materials in the Couette flow with a low Reynolds number. 

 However, we should note that the mean value of such instantly changing ensemble configurations is close to the exact solution, as shown in Figs. \ref{fig-4}(a) and (b). The mean values of $\vec{V}$ are graphically shown in Fig. \ref{fig-8} (c) for $D_{\rm sim}\!=\!1$ and Fig. \ref{fig-8} (d) for $D_{\rm sim}\!=\!1000$, both of which are close to the exact solution. The pressure $p$ is also the mean value in both Figs. \ref{fig-8}(c) and (d). Interestingly, the patterns of $p$ are almost the same though the magnitudes of $p$ are different to each other. This implies that the normalized  $p$ is independent of $D_{\rm sim}$ and is dependent only on the random numbers $\vec{g}_{ij}$  if the other simulation parameters are the same. Indeed, the patterns of $p$ in  Figs. \ref{fig-8} (a),(b) are almost the same, where the same random number sequence is used. The total number of convergent configurations is $1\times10^4$, as mentioned in the text. Therefore, the same random numbers are used for all simulations if the lattice size is the same. This is the reason why the pressure patterns in both Figs. \ref{fig-8}(c) and (d) are almost identical to each other.

\subsection{Results of dimensional analysis}
Now, we present three examples  of application of the statement (B) in Section \ref{dim-analysis}. These examples logically clarify responses of experimental results against the changes in physical parameters $\nu_e, V_e$ and $d_e$. The data (i), (ii) and (iii) are shown in Table \ref{table-5}. 

\begin{table}[!ht]
\centering
\caption{Three possible examples for modification of physical parameters $\nu_e, V_e$ and $d_e$ maintaining the condition $\gamma/\gamma_0\!=\!1$,  and the results obtained by the statement (B). }
\begin{tabular}{|c|c|c|c|c|c|c|c|c|c|c|c|c|c|}
\hline
\multicolumn{7}{|c|}{\bf Inputs}&\multicolumn{7}{|c|}{\bf Outputs} \\ \hline
& $\frac{\nu_e}{\nu_{e,0}}$ & $\frac{V_e}{V_{e,0}}$ &   $\frac{d_e}{d_{e,0}}$ & $\frac{\rho_e}{\rho_{e,0}}$  & $\frac{\tau_e}{\tau_{e,0}}$  & $\frac{D_e}{D_{e,0}}$& $\frac{\alpha}{\alpha_{0}}$ & $\frac{\beta}{\beta_{0}}$ & $\frac{\gamma}{\gamma_{0}}$ & $\frac{\delta}{\delta_{0}}$& $\frac{\lambda}{\lambda_{0}}$& $\textcolor{red}{\frac{\nu_{\rm sim}}{\nu_{0}}}$& $\textcolor{red}{\frac{D_{\rm sim}}{D_{0}}}$\\ \hline
(i)&2 &2&1&1&$\frac{1}{2}$&2&1&$\frac{1}{2}$&1&1&1&\textcolor{red}{1}&\textcolor{red}{$\frac{1}{4}$}\\ \hline
(ii)&1 &$\frac{1}{2}$&2&1&4&$\frac{1}{4}$&2&4&1&1&8&\textcolor{red}{1}&\textcolor{red}{4}\\ \hline
(iii)&2 &1&2&1&2&$\frac{1}{2}$&2&2&1&1&8&\textcolor{red}{1}&\textcolor{red}{1}\\ \hline
\end{tabular}
\label{table-4}
\end{table}
From these data ${\nu_e}/{\nu_{e,0}}$, ${V_e}/{V_{e,0}}$,   ${d_e}/{d_{e,0}}$ and ${\rho_e}/{\rho_{e,0}}$, and Eqs. (\ref{param-ratio-1}), (\ref{existence-parameters-1}), (\ref{existence-parameters-2}) in Appendix \ref{APP-B}, we obtain ${\tau_e}/{\tau_{e,0}}, ... , {\lambda}/{\lambda_0}$  in Table  \ref{table-4}. The density $\rho_e$ is assumed to be the same as $\rho_{e,0}$, as mentioned above, and hence, ${\rho_e}/{\rho_{e,0}}\!=\!1$. Thus, from Eq. (\ref{existence-parameters-3}) and the relations 
$\nu_{e,0}=\alpha_0^2\beta_0\gamma_0^{-1}\nu_0,\quad D_{e,0}={\alpha_0^{-2}{\beta_0^3\gamma_0^2}\delta_0^{-1}}D_0$, we obtain
\begin{eqnarray}
\label{final-results}
\begin{split}
&\frac{\nu_{\rm sim}}{\nu_0}=\left(\frac{\alpha}{\alpha_0}\right)^{-2}\left(\frac{\beta}{\beta_0}\right)\left(\frac{\gamma}{\gamma_0}\right)^{-1}\frac{\nu_e}{\nu_{e,0}},\\
&\frac{D_{\rm sim}}{D_0}=\left(\frac{\alpha}{\alpha_0}\right)^{-2}\left(\frac{\beta}{\beta_0}\right)^{3}\left(\frac{\gamma}{\gamma_0}\right)^{2}\left(\frac{\delta}{\delta_0}\right)^{-1}\frac{D_e}{D_{e,0}}.
\end{split}
\end{eqnarray}
Using these formulas, we have the data in the final two columns (red colored letters) in Table \ref{table-4}. The values of ${D_{\rm sim}}/{D_0}$ and ${\nu_{\rm sim}}/{\nu_0}$ can also be obtained from the model in Ref.~\cite{Egorov-etal-POF2020} because the scaling relations for $\nu$ for $\gamma\!=\!1$ and $D$ in Eq. (\ref{parameter-scale-1}) are the same as in Ref.~\cite{Egorov-etal-POF2020}.

\begin{figure}[!h]
\centering{}\includegraphics[width=10.0cm]{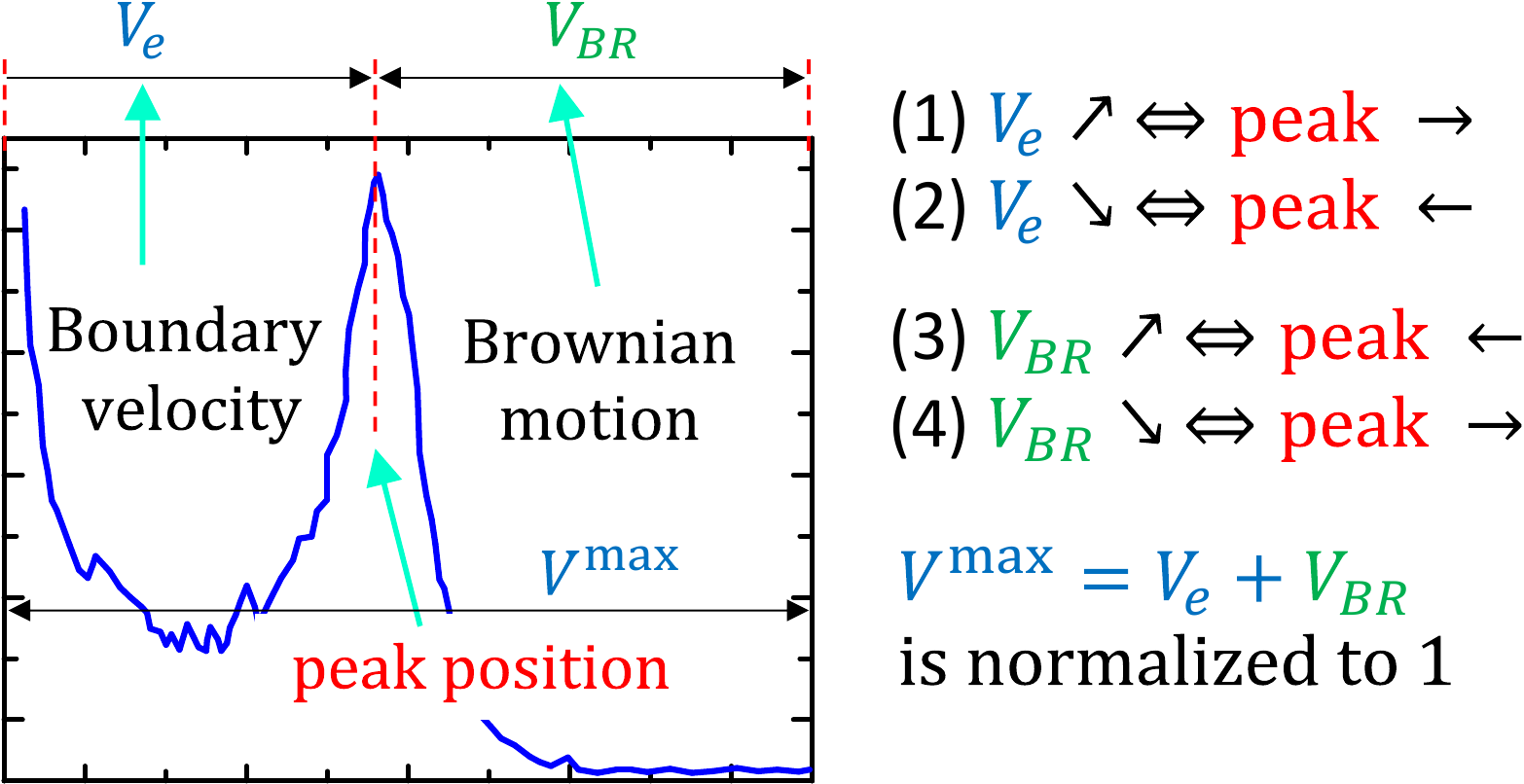}
\caption{An illustration of experimentally observed peak position and its relation to  the boundary velocity $V_e$ (or $V_B$) and velocity $V_{BR}$ from Brownian motion.  (1) ``$V_e \nearrow\; \Leftrightarrow {\rm peak} \rightarrow$'' means that an increment of $V_e$ equivalent with a right move of the peak position,  (2) ``$V_e \searrow\; \Leftrightarrow {\rm peak} \leftarrow$'' means that a decrement of $V_e$ equivalent with a left move of the peak position. In these statements, $V_e$ can be replaced by  $V_{BR}$  if ``$\rightarrow$'' is replaced by ``$\leftarrow$'', and we obtain the statements (3) and (4). Note that $V^{\rm max}\!=\!V_{e}\!+\!V_{BR}$ is normalized to $V^{\rm max}\!=\!1$, which is the reason for the displacements of the peak position  explained by the statements (1)-(4). }
\label{fig-9}
\end{figure}

Here, we intuitively discuss whether the results in Table \ref{table-4} are physically meaningful and suggestive. First, we should note that the second peak in $h(V_x)$ and  $h(V)$  in Figs. \ref{fig-4}, \ref{fig-6} and \ref{fig-7}  correspond to the boundary velocity $V_B$, which is denoted by $V_{e}$. Note also that the maximum velocity $V_x^{\rm max}$ corresponds to $V_B$ plus the maximum Brownian speed in the $x$ direction, where the maximum Brownian speed is independent of $V_B$, as discussed in Ref.~\cite{Egorov-etal-POF2020}. The reason why the peak position changes according to the $D_{\rm sim}$ variation is that $V_x^{\rm max}$ increases (decreases) with increasing (decreasing) $D_{\rm sim}$ and is normalized to $V_x^{\rm max}\!=\!1$, while $V_B$ remains fixed. These relations are intuitively illustrated in Fig. \ref{fig-9}, where the symbol $V_{BR}$ denotes velocity or velocity increment by Brownian motion of fluid particles.

Thus, result (i) in Table \ref{table-4} is suggestive. If the kinematic viscosity and the boundary velocity are increased such that $\nu_e/\nu_{e,0}\!=\!2$ and $V_e/V_{e,0}\!=\!2$, then we have the result $D_{\rm sim}/D_0\!=\!1/4$. This result indicates that the peak position moves right, which implies that $V_x^{\rm max}$ decreases because velocity of the Brownian motion reduces as indicated in Fig. \ref{fig-9}. Moreover, $\nu_{\rm sim}$ remains unchanged due to the assumption that the lattice size remains unchanged, and therefore, $\nu_{\rm sim}$ does not affect the peak position. We should note that these increments of  $\nu_e/\nu_{e,0}$  and $V_e/V_{e,0}$ make no change to the Reynolds number.  On the other hand,  $V_e/V_{e,0}\!=\!2$  moves the peak right because the boundary velocity corresponds the peak position, and $D_e/D_{e,0}\!=\!2$ implies an increment of speed of water molecules, and hence, the maximum speed $V_x^{\rm max}$ is expected to increase implying that the peak moves left. 
Therefore, it is unclear whether the peak position moves right or left in $h(V_x)$, where  $V_x^{\rm max}$ is normalized. Thus, the result that $D_{\rm sim}/D_0\!=\!1/4$, implying that the peak position moves right, is non-trivial. 

The second result (ii) indicates that the system, in which the boundary velocity is $V_B/2$ in $E_e$, can be simulated by  $\nu_{\rm sim}\!=\!\nu_0$ and $D_{\rm sim}\!=\!4D_0$.  Therefore, the peak position is expected to move left.  On the other hand, the condition $V_e/V_{e,0}\!=\!1/2$ moves the peak position left, while $D_e/D_{e,0}\!=\!1/4$ decreases $V_x^{\rm max}$ moving the peak right. Therefore, it is also unclear whether the peak position moves right or left in the normalized $h(V_x)$, and hence, the result is also non-trivial.

The third result (iii) that $\nu_{\rm sim}\!=\!\nu_0$ and $D_{\rm sim}\!=\!D_0$ indicates that the peak position remains the same as in the original system. This result is also non-trivial for the same reasons as for results (i) and (ii).

These results are obtained by assuming $\gamma/\gamma_0\!=\!1$, and moreover the scaling relation for the simulation parameters are the same as in the model in Ref.~\cite{Egorov-etal-POF2020}. Thus, we find that the results obtained by the dimensional analysis in this paper are consistent with those obtained in Ref.~\cite{Egorov-etal-POF2020}

\section{Conclusion}
We study the velocity distribution of protoplasmic streaming in plant cells by two-dimensional (2D) Langevin Navier-Stokes (LNS) simulation for velocity and pressure, which is denoted by LNS($V,p$). In this study, the streaming is modified to 2D Couette flow in parallel plates with Brownian random force as in Ref.~\cite{Egorov-etal-POF2020}, where the LNS equation for vorticity and stream function, denoted by LNS($\psi,\omega$), is simulated. Since LNS($\psi,\omega$) is obtained by differentiating LNS($V,p$), a solution to LNS($V,p$) also satisfies LNS($\psi,\omega$). However, the converse statement is not always true. In other words, the reported numerical data obtained by LNS($\psi,\omega$) are not always reproduced by LNS($V,p$). This implies that we need to check whether numerical solutions of LNS($V,p$) are consistent with those of LNS($\psi,\omega$) or with the reported experimental data. This is why we apply LNS($V,p$) to simulate the 2D Couette flow in this paper. 

The simulation results of LNS($V,p$) are slightly different from those of LNS($\psi,\omega$) in Ref.~\cite{Egorov-etal-POF2020} in the sense that the shape of velocity distributions $h(V_x)$ and $h(V)$ are not exactly identical with those reported in Ref.~\cite{Egorov-etal-POF2020}. However, this slight deviation is not a failure in the simulation techniques but is considered to be a possible discrepancy, as mentioned above. The critical point is that the experimentally observed fact that there appear two different peaks in $h(V)$ at $V\!\to\!0$ and $|V_x|\!\not=\!0$ is reproduced in the simulations. Indeed,  two different peaks in $h(V)$ are apparently reproduced by the numerical solution of LNS($V,p$). 
 Moreover, we obtain valuable information on the solution under a scale transformation, including unit changes, by dimensional analysis of LNS($V,p$) like in the case of LNS($\psi,\omega$) in Ref.~\cite{Egorov-etal-POF2020}. The invariant property of LNS($V,p$) under the scale transformation is also slightly different from that of LNS($\psi,\omega$). Nevertheless, the responses of the solution of LNS($V,p$) reflected in the peak position at $|V|\!\not=\!0$ in $h(V_x)$ under a change of parameters, such as kinematic viscosity, system size, and boundary velocity, are compatible with the case of LNS($\psi,\omega$).

To summarize, we successfully simulate two different and experimentally observed peaks in the velocity distribution of the protoplasmic streaming by 2D LNS equation for velocity and pressure. From the results reported in this paper, we expect that 3D LNS simulations would be meaningful and feasible for a more detailed study of the flow field of protoplasmic streaming.

\section*{Acknowledgments}
The author H.K. acknowledges Andrey Shobukhov for the helpful discussions. 
This work is supported in part by a Collaborative Research Project J20Ly09 of the Institute of Fluid Science (IFS), Tohoku University, and in part by a Collaborative Research Project of the National Institute of Technology (KOSEN), Sendai College.  Numerical simulations were performed on the Supercomputer system "AFI-NITY" at the Advanced Fluid Information Research Center, Institute of Fluid Science, Tohoku University.

\appendix

\section{Proof of Eq. (\ref{parameter-scale-1}) \label{APP-A} } 
Using the relations $\vec{V}(\frac{\rm m}{\rm s})\!=\! \alpha^{-1}\beta \vec{V}(\frac{\rm \alpha m}{\rm \beta s})$, $\nu(\frac{\rm m^2}{\rm s})\!=\! \alpha^{-2}\beta\nu(\frac{\rm \alpha^2 m^2}{\rm \beta s})$, $\rho(\frac{\rm kg}{\rm m^3})\!=\! \alpha^3\lambda^{-1}\rho(\frac{\rm \lambda  kg}{\rm \alpha^3 m^3})$,  $p(\frac{\rm kg}{\rm m s^2})\!=\! \alpha \beta^2\lambda^{-1}p(\frac{\rm \lambda  kg}{\rm \alpha\beta^2 m s^2})$,  $D(\frac{\rm m^2}{\rm s^3})\!=\! \alpha^{-2} \beta^3D(\frac{\rm\alpha^2 m^2}{\rm\beta^3 s^3})$ and the transformations ${\it \Delta} t({\rm s})\!=\!\frac{\tau}{n_T}({\rm s})\!\to\!\frac{\beta^{-1} \tau}{\delta n_T}({\rm\beta s})\!=\!  \beta^{-1} \delta^{-1}{\it \Delta} t({\rm\beta s})$ and ${\it \Delta} x({\rm m})\!=\!\frac{d}{n_X}({\rm m})\!\to\!\frac{\alpha^{-1} d}{\gamma n_X}({\rm\alpha m})\!=\! \alpha^{-1} \gamma^{-1}{\it \Delta} x({\rm\alpha m})$, we obtain
\begin{eqnarray}
\label{NS-eq-time-step-transf-1}
\begin{split}
&{\vec V}(t+{\it \Delta} t) \\
&\leftarrow\vec{V}(t) +\delta^{-1}{\it \Delta} t \left[\gamma \left (-{\vec V}\cdot \nabla\right){\vec V}-\gamma{\rho}^{-1} {\it \nabla} p +\gamma^2\nu {\it \Delta} {\vec V}\right] +\sqrt{2D\delta^{-1}{\it \Delta} t} \,\vec{g},
\end{split}
\end{eqnarray}
where the common factor $\alpha^{-1}\beta$ is eliminated. From this, we obtain 
\begin{eqnarray}
\label{NS-eq-time-step-transf-2}
{\vec V}(t+{\it \Delta} t)\leftarrow \vec{V}(t) +\delta^{-1}\gamma{\it \Delta} t \left[ \left (-{\vec V}\cdot \nabla\right){\vec V}-{\rho}^{-1} {\it \nabla} p +\nu {\it \Delta} {\vec V}\right] +\delta^{-1}\gamma\sqrt{2D{\it \Delta} t} \,\vec{g},
\end{eqnarray}
if $\nu$ and $D$ scale according to $\nu \to \gamma^{-1}\nu$ and $D \to \gamma^2\delta^{-1}D$. However, $\nu$ is used for an additional assumption ($A\sim \tau/\nu$), which is explained in Appendix \ref{APP-B}. For this reason, $\gamma^{-1}\nu$ influences this assumption, and for this reason, we exclude the  transformation $n_T\!\to\!\gamma n_T$  by imposing the condition $\gamma\!=1\!$ for simplicity. Thus, 

\begin{eqnarray}
\label{parameter-scale-3}
\nu \to \gamma^{-1}\nu, \quad D \to \gamma^2\delta^{-1}D, \quad(\gamma=1).
\end{eqnarray}
For convergent solution ($\Leftrightarrow {\vec V}(t\!+\!{\it \Delta} t)\!=\! \vec{V}(t)$ at sufficiently large $t$, the factor $\delta^{-1}\gamma$ can be dropped. Thus, we find that the numerical solution of LNS equation in Eq. (\ref{NS-eq-org}) is invariant under the scale transformation in Eq. (\ref{scale-transf}) together with that in Eq. (\ref{parameter-scale-3}).

\section{ Proof of the statement (B). \label{APP-B} }
The outline of the proof of the statement (B) is as follows: 
First, we assume that the macroscopic relaxation time $\tau_e$ is given by
\begin{eqnarray}
\label{macro-relation}
\tau_e\simeq A_e/\nu_e,
\end{eqnarray}
where the area $A_e$ is written by using the diameter  $d_e$ such that $A_e\simeq d_e^2$ \cite{Zaichik-etal-ExpTFS1997}. Thus, from the Einstein-Stokes-Sutherland formula $D_{\rm diff}=k_BT/6\pi\mu a$ and the relation $2D_e\tau_e^2=D_{\rm diff}$, we have  $D_e\simeq \frac{k_BT}{\mu a}\tau_e^{-2}\simeq \frac{k_BT}{\mu a}(\frac{\nu_2}{A_e})^2\simeq \nu_ed_e^{-2}$, where $k_B$ is the Boltzmann constant and $T$ is the temperature,  $\mu$ and $a$ are viscosity and particle size, respectively~\cite{Egorov-etal-POF2020}. Thus, we have
\begin{eqnarray}
\label{param-ratio-1}
\frac{\tau_e}{\tau_{e,0}}=\frac{\nu_{e,0}}{\nu_e}\left(\frac{d_e}{d_{e,0}}\right)^2,\quad \frac{D_e}{D_{e,0}}=\frac{\nu_{e}}{\nu_{e,0}}\left(\frac{d_e}{d_{e,0}}\right)^{-2}
\end{eqnarray}
From the unit changes between $E_{e}=(\rho_e,\nu_e,V_e,d_e)$ and $(\rho_0,\nu_0,V_0,d_0)(\subset S)$, and between $E_{e,0}=(\rho_{e,0},\nu_{e,0},V_{e,0},d_{e,0})$ and $(\rho_0,\nu_0,V_0,d_0)$, we have $\alpha\!=\!\frac{\nu_eV_0}{\nu_0V_e}$, $\alpha_0\!=\!\frac{\nu_{e,0}V_0}{\nu_0V_{e,0}}$, $\beta\!=\!\frac{\nu_e}{\nu_0}\left(\frac{V_0}{V_e}\right)^2$, $\beta_0\!=\!\frac{\nu_{e,0}}{\nu_0}(\frac{V_0}{V_{e,0}})^2$, $\lambda\!=\!\frac{\rho_{e}}{\rho_0}(\frac{\nu_{e}V_0}{\nu_0V_{e}})^3$, and $\lambda_0\!=\!\frac{\rho_{e,0}}{\rho_0}(\frac{\nu_{e,0}V_0}{\nu_0V_{e,0}})^3$. Thus, we obtain
\begin{eqnarray}
\label{existence-parameters-1}
\frac{\alpha}{\alpha_0}=\frac{\nu_eV_{e,0}}{\nu_{e,0}V_e},\quad \frac{\beta}{\beta_0}=\frac{\nu_e}{\nu_{e,0}}\left(\frac{V_{e,0}}{V_e}\right)^2,\quad \frac{\lambda}{\lambda_0}=\frac{\rho_e}{\rho_{e,0}}\left(\frac{\nu_eV_{e,0}}{\nu_{e,0}V_e}\right)^3,
\end{eqnarray}
and
\begin{eqnarray}
\label{existence-parameters-2}
\frac{\gamma}{\gamma_0}(=1)=\frac{d_e\nu_{e,0}V_e}{d_{e,0}\nu_eV_{e,0}},\quad \frac{\delta}{\delta_0}=\frac{\tau_e\nu_{e,0}}{\tau_{e,0}\nu_{e}}\left(\frac{V_{e}}{V_{e,0}}\right)^2
\end{eqnarray}
from the relations in Eq. (\ref{dt-dx}). Using these parameters $\alpha,\beta,\gamma,\lambda,\delta$ in Eqs. (\ref{existence-parameters-1}), (\ref{existence-parameters-2}), we define
\begin{eqnarray}
\label{existence-parameters-3}
\nu_{\rm sim}=\alpha^2\beta\gamma^{-1}\nu_e,\quad D_{\rm sim}={\alpha^{-2}{\beta^3\gamma^2}\delta^{-1}}D_e, \quad(\gamma=1).
\end{eqnarray}
The lattice spacing ${\it \Delta}x_0$ and time step ${\it \Delta}t_0$ are given by Eq. (\ref{dt-dx}) such that
\begin{eqnarray}
\label{dx-dt-scale-trans}
\begin{split}
&{\it \Delta}x_0=\frac{d_e}{\gamma n_X}\alpha^{-1}=\frac{d_e}{\gamma n_X}\frac{\nu_0}{\nu_e}\frac{V_e}{V_0} \;(\alpha {\rm m}), \quad(\gamma=1), \\
&{\it \Delta}t_0=\frac{\tau_e}{\delta n_T}\beta^{-1}=\frac{\tau_e}{\delta n_T}\frac{\nu_0}{\nu_e}\left(\frac{V_e}{V_0}\right)^2 \;(\beta {\rm s}), 
\end{split}
\end{eqnarray}
where $d_e/n_X$ and $\tau_e/n_T$ can be written as ${\it \Delta}x_e\!=\!d_e/n_X$ and ${\it \Delta}t_e\!=\!\tau_e/n_T$, and therefore we have
\begin{eqnarray}
\label{dx-dt-real-def}
{\it \Delta}x_e=\gamma \alpha{\it \Delta}x_0, \quad
{\it \Delta}t_e=\delta\beta{\it \Delta}t_0, \quad(\gamma=1).
\end{eqnarray}
Thus, we obtain 
\begin{eqnarray}
\label{proof-1}
\begin{split}
&\left(\rho_0, \nu_{\rm sim},V_{B0}, D_{\rm sim}, {\it \Delta}x_0, {\it \Delta}t_0\right) \\
=&\left(\frac{\alpha^3}{ \lambda}\rho_{e}, \frac{\beta}{\alpha^{2}\gamma}\nu_e,\frac{\beta}{\alpha} V_{Be}, \frac{\beta^3\gamma^2}{\alpha^{2}\delta}D_e, \frac{1}{\alpha\gamma}{\it \Delta}x_e,\frac{1}{\beta\delta} {\it \Delta}t_e\right), \quad(\gamma=1),
\end{split}
\end{eqnarray}
which implies that ${\rm Exp}(E_{e})$ can be simulated with $(\rho_0, \nu_{\rm sim},V_{B0}, D_{\rm sim}, {\it \Delta}x_0, {\it \Delta}t_0)$.


%
%
%

\end{document}